\documentclass[letter]{aa}
\usepackage{natbib}
\usepackage{graphicx}
\usepackage{txfonts}
\usepackage{hyperref}
\begin{document}

\title{X-SHOOTER Spectrum of Comet 3I/ATLAS: Insights into a Distant Interstellar Visitor}

\author{
A. Alvarez-Candal%\inst{1}
\and
J.L. Rizos%\inst{1}
\and
L.M. Lara%\inst{1}
\and
P. Santos-Sanz%\inst{1}
\and
P.J. Gutierrez%\inst{1}
\and
J.L. Ortiz%\inst{1}
\and
N. Morales%\inst{1}
%\and
%J. de León\inst{2,3}
}

\institute{
%1
Instituto de Astrofísica de Andalucía – Consejo Superior de Investigaciones Científicas (IAA-CSIC), Glorieta de la Astronomía S/N, E-18008, Granada, Spain
%\and
%2
%Instituto de Astrofísica de Canarias (IAC), C/Vía Láctea S/N, E-38205, La Laguna, Spain
%\and
%3
%Departamento de Astrofísica (ULL), E-38205, La Laguna, Spain
}

\date{Received \today; accepted XXX, 2025}

% \abstract{}{}{}{}{} 
% 5 {} token are mandatory
 
\abstract
  % context heading (optional)
  % {} leave it empty if necessary  
   {Comets are primitive remnants of the early Solar System whose composition offers fundamental clues to their formation and evolution. High-resolution, broad-wavelength spectroscopy is crucial for identifying volatile species and constraining the physical conditions within the coma.}
  % aims heading (mandatory)
   {We aim to characterize the gas composition and physical environment of the newly discovered comet 3I/ATLAS through optical and near-infrared spectroscopy.}
  % methods heading (mandatory)
   {We used a medium-resolution spectrum of comet 3I/ATLAS with X-shooter at the ESO Very Large Telescope (VLT), covering the 300–2500 nm wavelength range. Standard data reduction and flux calibration were applied.}
  % results heading (mandatory)
   {Although the object clearly shows activity, only upper limits to the production rates of OH and CN can be estimated: $8.2\times10^{26}$ s$^{-1}$ and $5.6\times10^{23}$ s$^{-1}$, respectively. We obtained red spectral slopes consistent with those of typical D-type asteroids and objects in the outer Solar System.}% It corresponds to a spectral slope at least a factor of $\sim1.86$ higher than that of 2I/Borisov. }
  % conclusions heading (optional), leave it empty if necessary 
   {}


   \maketitle
%
%-------------------------------------------------------------------

\section{Introduction}

Interstellar objects are defined as bodies not gravitationally bound to the Sun. The first such object discovered passing through our Solar System was 1I/'Oumuamua (1I/2017 U1), detected in 2017 by Robert Weryk using the Pan-STARRS telescope at Haleakalā Observatory, in Hawaii. 'Oumuamua is a small object exhibiting highly variable light curves, suggesting an extremely elongated shape \citep{Jewitt2017}—possibly more so than any known Solar System object. It displays a reddish color \citep{Ye2017}, similar to bodies in the outer Solar System. Despite its close perihelion passage, 'Oumuamua showed no detectable coma \citep{Jewitt2017}, although a non-gravitational acceleration was measured, possibly resulting from outgassing or solar radiation pressure \citep{Micheli2018}. 

The second confirmed interstellar object was 2I/Borisov, originally designated C/2019 Q4 (Borisov). 2I/Borisov was discovered on August 29, 2019, by Gennadiy Borisov in Crimea. Unlike 'Oumuamua, 2I/Borisov showed clear cometary activity \citep{Opitom2019}, with a visible coma and tail. Its perihelion was 2.006 AU, and it had an eccentricity of 3.357. Its estimated diameter is $<0.5$ km, assuming a typical cometary albedo \citep{Jewitt2019,Guzik2020}.

A new interstellar visitor, 3I/ATLAS (and provisionally designated A11pl3Z), was discovered on July 1, 2025, by the ATLAS survey at a heliocentric distance of 4.53 AU. Precovery observations from June 14 reveal an extremely high eccentricity ($\sim$6.0), nearly twice that of 2I / Borisov, the most eccentric interstellar object known to date, strongly supporting its extrasolar origin \citep{seligman2025arXiv}. {The object was bright enough (V$\approx18.1$) and displays signs of cometary activity, including a detectable coma \citep{jewitt2025ATel,alarcon2025ATel}.}

Understanding the surface composition of interstellar objects (ISOs) is essential for constraining their thermal and dynamical histories, as well as their potential connections to small body populations in other planetary systems \citep{Seligman2022}. These objects may preserve primitive materials that formed around other stars, providing a rare opportunity to probe the diversity of planetary formation environments beyond our own Solar System. 

Spectroscopy across a broad wavelength range—from the near-ultraviolet to the near-infrared—makes it possible to detect absorption features associated with specific ices (e.g., H$_2$O), organic compounds, and surface alteration processes \citep{Jewitt2024}. It also provides insight into the effects of prolonged exposure to interstellar radiation and cosmic rays, as well as early signs of activity induced by solar heating \citep{Guzik2020}. By comparing the spectral properties of ISOs with those of comets, asteroids, and trans-Neptunian objects in the Solar System, we can investigate whether similar chemical and physical processes operate universally or differ markedly across stellar systems.

Since its discovery, 3I/ATLAS has gathered widespread attention from the community. \cite{seligman2025arXiv} showed a red photo-spectrum, with a spectral slope ($S'$) about 20 \%/1000 \AA\ (visually estimated from their Fig. 6), with an estimated magnitude variation of less than 0.2 mag on a time-span of 29 h, and an estimated diameter of about 10 km. Interestingly, \cite{bolin2025arXiv} measured solar colors (or an almost neutral $S'$) while \cite{Opitom2025} obtained a spectrum with a red slope, of the order of 20\%/ 1000 \AA. \cite{hopkins2025arxiv} suggests that 3I/ATLAS may have originated within the thick disk of the Milky Way.

In this work, we present the spectroscopic observation of comet 3I/ATLAS obtained with the X-SHOOTER instrument mounted on ESO’s Very Large Telescope (VLT). This dataset provides a unique, simultaneous view across the ultraviolet, visible, and near-infrared spectral domains. These observations represent a significant step forward in characterizing dynamically new comets. Acquiring spectra at this early stage offers a valuable baseline before the onset of intense solar-driven evolution, enhancing the scientific relevance of the dataset. This work is organized as follows: Section \ref{sec:data_reduction} describes the data reduction, Section \ref{sec:res_dis} presents the analysis and discussion of the results. The last Section presents the Conclusions of this work.

\section{Observations and data reduction}\label{sec:data_reduction}

We used the spectra of the interstellar object 3I/ATLAS, obtained on two nights: July 4, 2025, between 06:02 and 07:00 UT, and July 5, between 01:50 and 02:27 UT. The instrument used was X-SHOOTER mounted on Unit 3 of the VLT. The observations were conducted in service mode under DDT program 115.29F3.001 (PI: Puzia), the data being non-proprietary and publicly available to the scientific community. The spectrograph simultaneously covers the full 0.3–2.5~$\mu$m spectral range by splitting the incoming light from the telescope into three separate arms using two dichroic filters. The instrument setup was
$$
\begin{array}{c | c | c | c}
       {\rm ARM}  & {\rm Slit} & {\rm Readout\ mode} & {\rm EXPTIME\ (sec)} \\
       UVB        &  1.6\times11 & {\rm 100k,\ high\ gain} & 900\\
       VIS        &  1.5\times11 & {\rm 100k,\ high\ gain} & 963\\
       NIR        &  1.2\times11 & {\rm Default}           & 300\times3.\\
\end{array}$$
Observations were done using generic offsets to facilitate subsequent sky signal estimation and subtraction. A detailed description of the instrument is available at \cite{Vernet2011}. In total, six spectra were obtained (three per night). At the time of observation, there were excellent visibility conditions (seeing lower than 1\arcsec) and photometric sky conditions. {During the first night, the acquired spectrum captured only light from 3I/ATLAS since the object was rather isolated in the field; however, on the second night, the field was more crowded, but not hampering our efforts to acquire the long-slit spectrum. Nevertheless, the acquisition images show that the object was free from obvious contamination by background objects.}

%\subsection{Data reduction}

To reduce the X-SHOOTER spectra, we first processed the data through the ESO esoreflex pipeline version 2.1.9\footnote{\url{https://www.eso.org/sci/software/esoreflex/}}. The pipeline removes the instrument signature, i.e., debiases, flat-fields, wavelength-calibrates, orders-merges, extracts, sky-subtracts, and finally flux-calibrates the data. This renders 2D spectra of 3I/ATLAS, which are combined to increase the signal-to-noise ratio (S/N) after centering them on the comet nucleus (i.e., the optocenter). The whole process relies on daily and ancillary calibration files. We used our tools to extract the final 1D spectra from the 2D images.

The observational strategy did not include observing a solar analogue star to remove the solar signature. Thus, {to analyze the refractory component of the coma} we used the solar spectrum provided by the LISIRD database\footnote{\url{https://lasp.colorado.edu/lisird/data/tsis_ssi_24hr/}}, based on the TSIS-1 SIM instrument. We retrieved solar irradiance data for the week of June 7–13, 2025, which was the most recent available at the time of our analysis. This period is particularly relevant, as the Sun is currently near its activity maximum (maximum of Solar Cycle 25, which began in December 2019), making it essential to use solar conditions that closely match those during the observation.

Then, we averaged these daily solar spectral irradiance values (in W\,m$^{-2}$\,nm$^{-1}$) over these seven days and converted the result to the solar flux as observed at 4.37 au, which was the heliocentric distance of the comet during the acquisition of its spectrum.

To reduce noise in the comet spectrum, we applied binning of {60 pixels}, equivalent to 1.2\,nm intervals. We then used the same bin width as the solar spectrum. To perform the division, we interpolated the binned solar irradiance onto the wavelength grid of the comet spectrum. Finally, we normalize the divided spectrum at 550 nm.

3I/ATLAS was also observed (with DDT) on 2025 July 3 from Calar Alto Observatory (Almería, Spain) using the 2.2-m telescope equipped with the CAFOS instrument (pixel scale of 0.53"/pixel). On that night, several images were acquired with the B, V, and R filters, each with an exposure time of 60 s. Although nearby stars probably contaminated the signal of the comet in many of the images, some of them allowed us to extract the uncontaminated brightness. Flux was calibrated with PanSTARRS DR2 catalog by using the transformations from \cite{2012ApJ...750...99T}.

\section{Results and Discussion}\label{sec:res_dis}

\begin{figure}[h]
\centering
\includegraphics[width=0.9\columnwidth]{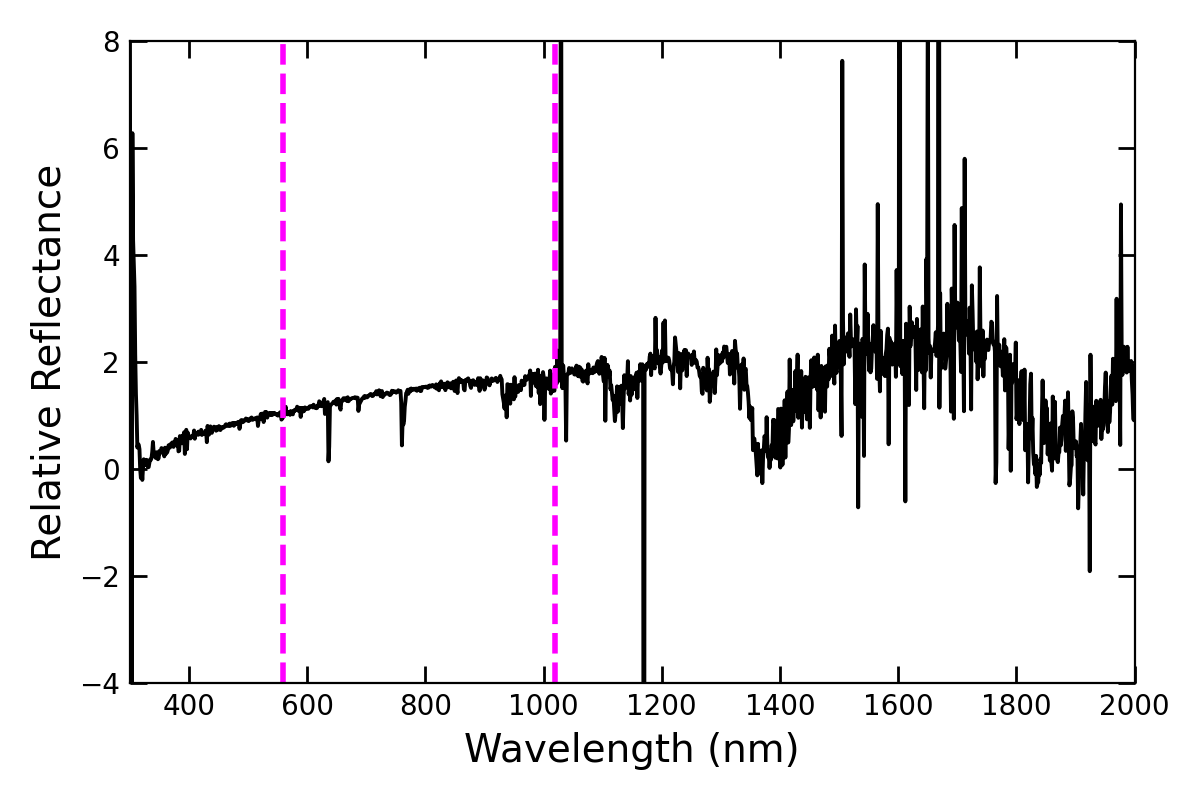}
\caption{Combined X-shooter reflectance spectrum of 3I/ATLAS, normalized at 550\,nm. The spectrum has been binned into 1.2\,nm intervals and divided by the solar flux. The pink vertical bars mark the limits of the UBV-VIS and VIS-NIR arms. The regions centered at 1400 and 1900\,nm are strongly affected by telluric absorption features.}
\label{fig:spectra}
\end{figure}

Figure \ref{fig:spectra} shows the combined X-SHOOTER reflectance spectrum of  3I/ATLAS, normalized at 550 nm and corrected for the solar contribution. The spectral slopes ($S'$) in the UVB and VIS ranges are:
\begin{itemize}
    \item UBV (320–560\,nm): 38.3~$\pm$~0.9\,\%/1000\,\AA
    \item VIS (560–1020\,nm): 12.5~$\pm$~0.6\,\%/1000\,\AA.
\end{itemize}

From our broadband photometry obtained at the Calar Alto observatory, it has been possible to estimate colors. We derive B-V = 1.07 $\pm$ 0.10 and V-R = 0.51 $\pm$ 0.06, values consistent with the slopes obtained from the spectrum, and also with those reported by \cite{Opitom2025} at 1-$\sigma$ level.% and with the values reported above.

The signal in the NIR range is strongly affected by telluric absorption, making the data unreliable for computing $S'$ in that region.

We used the X-SHOOTER spectrum to compute $S'$ in the same visible spectral ranges as in  \cite{Opitom2025} from data acquired with VLT/MUSES. The results are shown below
\begin{itemize}
    \item 500–700\,nm: 20.2~$\pm$~1.6\,\%/1000\,\AA
    \item 700–900\,nm: 15.9~$\pm$~1.7\,\%/1000\,\AA
    \item 500–900\,nm: 18.3~$\pm$~0.6\,\%/1000\,\AA,
\end{itemize}
being consistent with those reported by the authors of 18$\pm$3,  17$\pm$4, and 18$\pm$4 \,\%/1000\,\AA, respectively. In the spectral range from 528 to 860~nm, we measured a spectral slope for object 3I/ATLAS of $S' = 18.59 \pm 0.80~\%/1000~\text{\AA}$. These values are also in good agreement with the results reported by \cite{delafuente2025arXiv} using OSIRIS/GTC in La Palma, {\cite{seligman2025arXiv} using SNIFS at the UH 2-m telescope in Maunakea, and \cite{belya2025arXiv} using NGPS at Mount Palomar 200 in telescope}. This value indicates a significantly redder spectrum compared to comet 2I/Borisov, which exhibited a slope of $4$–$10~\%/1000~\text{\AA}$ over the same wavelength range \citep{deam2025}. This corresponds to a spectral slope of 3I/ATLAS at least a factor of $\sim1.86$ higher than that of 2I/Borisov, assuming its maximum reported value of $10~\%/1000~\text{\AA}$. A more pronounced reddening could indicate a different particle size distribution, possibly with a higher proportion of larger dust particles (which scatter red light more efficiently). {Alternatively, it may result from the surface material having been exposed to interstellar radiation and cosmic rays that altered the surface chemical composition for a long period during its interstellar travel, leading to more extensive space weathering that may result in redder colors \cite{brunetto2015aste.book,zhang2022}}. These discrepancies highlight the complexity of characterizing these distant interstellar visitors and underscore the need for continued, coordinated observations to understand their nature and origin fully.

\begin{figure}[h]
\centering
\includegraphics[width=0.9\columnwidth]{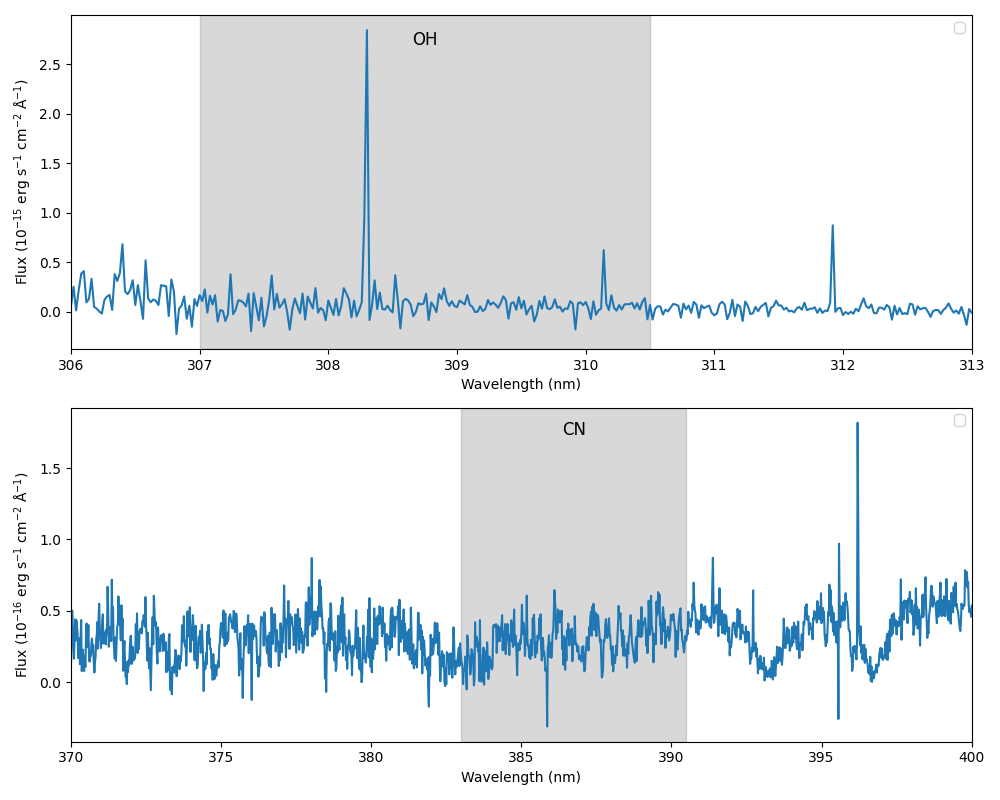}  
\caption{Spectrum of 3I/ATLAS{, after removing the solar signature, }showing the wavelength ranges where the gas emission bands should appear above the noise level. {Top: OH region. The three apparent emission lines are hot pixels remaining from the data reduction and not real emission lines within the OH band. Bottom: CN region and two adjacent continua.} }
\label{fig:gas}
\end{figure}

Figure \ref{fig:gas} shows a zoomed-in view of the spectrum, highlighting the OH and CN emission regions, specifically within the 306-313 and 370-400 nm intervals. The OH and CN bands should appear in the shaded areas if they existed. It cannot be firmly concluded that any of the gases are detected in the coma of 3I/ATLAS; thus, only upper limits to the production rates $Q$ in mol s$^{-1}$ can be derived. {To remove the solar signature, we have used a high-spectral resolution solar spectrum\footnote{\tt http://kurucz.harvard.edu/sun/irradiance2005/irradthu.dat} \citep{kuru2005MSAIS} scaled to the flux of our spectrum by matching H and K Ca II lines in both spectra}.  The fluorescence scattering efficiency (so-called $g$-factor) is a function of heliocentric velocity. For the comet velocity of -57.54 km/s and $\mathrm r_h=4.4$ AU at the time of our observations, the obtained $g$-factors are $6.25\times 10^{-15} r_h^{-2}$ and $3.34\times 10^{-13} r_h^{-2}$ erg/s/mol for OH ($ \Delta \nu=0$) and CN ($ \Delta \nu=0$) respectively \citep{Schleicher2010AJ}. As the gaseous emission is not detected, we compute the $3 \sigma$ of the flux within the ranges 307 and 310.5 nm and 383.0 and 390.5 nm, for OH and CN, respectively, to estimate the gas production of these species. In this way, we obtain between 307 and 310.5 nm and 383.0 and 390.5 nm, {respectively, $7.4 \times 10^{-16}$ erg cm$^{-2}$ s$^{-1}$ and $ 4.6  \times 10^{-17}$ erg cm$^{-2}$ s$^{-1}$. The upper limit to the total number of OH and CN molecules within the equivalent circular aperture radius (1.5\arcsec) of the slit used in our observations (1.2\arcsec width and 11\arcsec length) is $7.4  \times 10^{28}$ and $ 8.6 \times 10^{25}$.} The OH and CN production rate is then derived from a simple Haser model \citep{haser1957} with $l_p=2.4 \times 10^4$ and $l_d=1.6 \times 10^5$ km for OH and $l_p=1.3 \times 10^4$ and $l_d=2.2 \times 10^5$ km for CN as the effective scale lengths at $r_h = 1$ AU \citep{ahearn1995}. The velocity of the constant gas radial expansion of the coma is assumed to be $v=1$ km/s. {Therefore, we find $Q_{CN} <  5.6 \times 10^{23}$ s$^{-1}$ and $Q_{OH} < 8.2 \times 10^{26}$ s$^{-1}$. Assuming that OH is produced by ${\rm H}_2{\rm O}$ photodissociation with a branching ratio of 0.9, the water production rate is estimated below {$9.1 \times 10^{26}$ s$^{-1}$}. }

%The log [$\mathrm Q_{CN}/{\rm Q}_{OH}]$ is -1.2, considerably higher than the average value of -2.5 derived in \cite{ahearn1995} for a large set of comets. However, our result must be taken with caution, given the non-detection of the gaseous emissions. 

\section{Conclusions}

3I/ATLAS is the third known interstellar visitor to the Solar System. As soon as it was announced, it became the target of numerous observational campaigns using various telescopes, instruments, and observational techniques aimed at gaining a deeper understanding of this visitor, as mentioned above. In this work, we present spectroscopic observations of the interstellar comet 3I/ATLAS obtained with the X-SHOOTER instrument at the ESO VLT, covering the 300--2500~nm spectral range. The reflectance spectrum, normalized at 550~nm and corrected for the solar contribution, exhibits a moderately red slope in the visible, consistent with previous measurements of 3I/ATLAS by \cite{seligman2025arXiv} and \cite{Opitom2025}, and broadly comparable to the spectral behavior of Solar System comets, trans-Neptunian objects, and D-type asteroids—the reddest class known \citep{demeocarry2013}. Notably, \cite{bolin2025arXiv} obtained colors indicating a neutral spectral slope, results that are at odds with ours. The possible color variations warrant further observations, especially multi-filter or spectroscopic follow-up.

No clear gas emission bands (e.g., OH or CN) were detected in the near-UV range. From the absence of significant features in the observed spectrum, {we derived upper limits to the production rates of CN and OH using a Haser model \citep{haser1957}, obtaining $Q_{\mathrm{CN}} < 5.6 \times 10^{23}$~mol~s$^{-1}$ and $Q_{\mathrm{OH}} < 8.2 \times 10^{26}$~mol~s$^{-1}$.}

These observations provide a valuable baseline for monitoring the evolution of 3I/ATLAS as it approaches perihelion. The lack of strong gas emission features in the UV (OH and CN) suggests that sublimation of water and hydrogen cyanide ices is not yet taking place at this heliocentric distance. For comparison, CN detection for 2I/Borisov was performed when the object was at a distance of 2.66 AU \citep{fitzsimmons2019ApJ} and OH was detected at 2.38 AU \citep{mckay2020ApJ}, in both cases closer to the Sun than the distance at which the X-SHOOTER observations were obtained.

Continued observations in the coming months will be critical in assessing the evolution of its coma and better constraining the nature and origin of this interstellar visitor. Last, the attention gathered by 3I/ATLAS shows the interest of the community for these objects, reinforcing the need for coordinated efforts to observe and follow up on them, considering that Rubin Observatory's Legacy Survey for Space and Time may observe up to 70 such objects per year \citep{marceta2023PSJ}, one of them maybe becoming the target of ESA mission Comet Interceptor, planned to be launched by 2029.

\begin{acknowledgements}
We thank the ESO DG and DDT panel for awarding time to this project, and the service mode observer, telescope operator, and staff who enabled us to get the data. The same acknowledgment applies to the Calar Alto Observatory (CAHA) and its staff. A. Alvarez-Candal, J.L. Rizos, P. Santos-Sanz, J.L. Ortiz, P.J. Gutierrez, and L.M. Lara acknowledge financial support from the Severo Ochoa grant CEX2021-001131-S funded by MCIN/AEI/10.13039/501100011033. JLR, LML, PJG acknowledge financial support from grant PID2021-126365NB-C21. A. Alvarez-Candal acknowledges financial support from the project PID2023-153123NB-I00, funded by MCIN/AEI. P. Santos-Sanz acknowledges financial support by the Spanish grants PID2022-139555NB-I00 and PDC2022-133985-I00 funded by MCIN/AEI/10.13039/501100011033 and by the European Union "NextGenerationEU"/PRTR. The results presented in this document are based on data measured by the TSIS-1 Spectral Irradiance Monitor (SIM). These data are available from the TSIS-1 website at \url{https://lasp.colorado.edu/tsis/data/}. These data were accessed via the LASP Interactive Solar Irradiance Datacenter (LISIRD) (\url{https://lasp.colorado.edu/lisird/}).
Based on observations collected at the European Southern Observatory under ESO programme 115.29F3.001. This research is also based on observations collected at the Centro Astronómico Hispano en Andalucía (CAHA), jointly operated by the Instituto de Astrofísica de Andalucía (IAA-CSIC) and Junta de Andalucía under programme DDT.25B.343.
\end{acknowledgements}

\bibliography{main}

\begin{thebibliography}{29}
\expandafter\ifx\csname natexlab\endcsname\relax\def\natexlab#1{#1}\fi

\bibitem[{{A'Hearn} {et~al.}(1995){A'Hearn}, {Millis}, {Schleicher}, {Osip}, \& {Birch}}]{ahearn1995}
{A'Hearn}, M.~F., {Millis}, R.~C., {Schleicher}, D.~O., {Osip}, D.~J., \& {Birch}, P.~V. 1995, \icarus, 118, 223

\bibitem[{{Alarcon} {et~al.}(2025){Alarcon}, {Serra-Ricart}, {Licandro}, {Arencibia}, {Ruiz Cejudo}, \& {Trujillo}}]{alarcon2025ATel}
{Alarcon}, M.~R., {Serra-Ricart}, M., {Licandro}, J., {et~al.} 2025, The Astronomer's Telegram, 17264, 1

\bibitem[{{Belyakov} {et~al.}(2025){Belyakov}, {Fremling}, {Graham}, {Bolin}, {Kilic}, {Jewett}, {Lisse}, {Ingebretsen}, {Ryleigh Davis}, \& {Wong}}]{belya2025arXiv}
{Belyakov}, M., {Fremling}, C., {Graham}, M.~J., {et~al.} 2025, arXiv e-prints, arXiv:2507.11720

\bibitem[{{Bolin} {et~al.}(2025){Bolin}, {Belyakov}, {Fremling}, {Graham}, {Gray}, {Ingebretsen}, {Jewett}, {Kilic}, {Lisse}, {Roderick}, {Abdelaziz}, {Abron}, {Coughlin}, {Elhosseiny}, {Hsieh}, {Ma{\v{s}}ek}, {Molham}, {Takey}, {Noll}, \& {Wong}}]{bolin2025arXiv}
{Bolin}, B.~T., {Belyakov}, M., {Fremling}, C., {et~al.} 2025, arXiv e-prints, arXiv:2507.05252

\bibitem[{{Brunetto} {et~al.}(2015){Brunetto}, {Loeffler}, {Nesvorn{\'y}}, {Sasaki}, \& {Strazzulla}}]{brunetto2015aste.book}
{Brunetto}, R., {Loeffler}, M.~J., {Nesvorn{\'y}}, D., {Sasaki}, S., \& {Strazzulla}, G. 2015, in Asteroids IV, ed. P.~{Michel}, F.~E. {DeMeo}, \& W.~F. {Bottke}, 597--616

\bibitem[{{de la Fuente Marcos} {et~al.}(2025){de la Fuente Marcos}, {Licandro}, {Alarcon}, {Serra-Ricart}, {de Leon}, {de la Fuente Marcos}, {Lombardi}, {Tejero}, {Cabrera-Lavers}, {Guerra Arencibia}, \& {Ruiz Cejudo}}]{delafuente2025arXiv}
{de la Fuente Marcos}, R., {Licandro}, J., {Alarcon}, M.~R., {et~al.} 2025, arXiv e-prints, arXiv:2507.12922

\bibitem[{Deam {et~al.}(2025)Deam, Bannister, Opitom, Knight, Ridden-Harper, Seligman, Fitzsimmons, Guilbert-Lepoutre, Jehin, Jorda, Marsset, Moulane, Rousselot, Vernazza, \& Yang}]{deam2025}
Deam, S.~E., Bannister, M.~T., Opitom, C., {et~al.} 2025 [\eprint[arXiv]{2507.05051}]

\bibitem[{{DeMeo} \& {Carry}(2013)}]{demeocarry2013}
{DeMeo}, F.~E. \& {Carry}, B. 2013, \icarus, 226, 723

\bibitem[{{Fitzsimmons} {et~al.}(2019){Fitzsimmons}, {Hainaut}, {Meech}, {Jehin}, {Moulane}, {Opitom}, {Yang}, {Keane}, {Kleyna}, {Micheli}, \& {Snodgrass}}]{fitzsimmons2019ApJ}
{Fitzsimmons}, A., {Hainaut}, O., {Meech}, K.~J., {et~al.} 2019, \apjl, 885, L9

\bibitem[{{Guzik} {et~al.}(2020){Guzik}, {Drahus}, {Rusek}, {Waniak}, {Cannizzaro}, \& {Pastor-Marazuela}}]{Guzik2020}
{Guzik}, P., {Drahus}, M., {Rusek}, K., {et~al.} 2020, Nature Astronomy, 4, 53

\bibitem[{{Haser}(1957)}]{haser1957}
{Haser}, L. 1957, Bulletin de la Societe Royale des Sciences de Liege, 43, 740

\bibitem[{Hopkins {et~al.}(2025)Hopkins, Dorsey, Forbes, Bannister, Lintott, \& Leicester}]{hopkins2025arxiv}
Hopkins, M.~J., Dorsey, R.~C., Forbes, J.~C., {et~al.} 2025, From a Different Star: 3I/ATLAS in the context of the \={O}tautahi-Oxford interstellar object population model

\bibitem[{{Jewitt}(2024)}]{Jewitt2024}
{Jewitt}, D. 2024, arXiv e-prints, arXiv:2407.06475

\bibitem[{{Jewitt} \& {Luu}(2019)}]{Jewitt2019}
{Jewitt}, D. \& {Luu}, J. 2019, \apjl, 886, L29

\bibitem[{{Jewitt} \& {Luu}(2025)}]{jewitt2025ATel}
{Jewitt}, D. \& {Luu}, J. 2025, The Astronomer's Telegram, 17263, 1

\bibitem[{{Jewitt} {et~al.}(2017){Jewitt}, {Luu}, {Rajagopal}, {Kotulla}, {Ridgway}, {Liu}, \& {Augusteijn}}]{Jewitt2017}
{Jewitt}, D., {Luu}, J., {Rajagopal}, J., {et~al.} 2017, \apjl, 850, L36

\bibitem[{{Kurucz}(2005)}]{kuru2005MSAIS}
{Kurucz}, R.~L. 2005, Memorie della Societa Astronomica Italiana Supplementi, 8, 189

\bibitem[{{Mar{\v{c}}eta} \& {Seligman}(2023)}]{marceta2023PSJ}
{Mar{\v{c}}eta}, D. \& {Seligman}, D.~Z. 2023, PSJ, 4, 230

\bibitem[{{McKay} {et~al.}(2020){McKay}, {Cochran}, {Dello Russo}, \& {DiSanti}}]{mckay2020ApJ}
{McKay}, A.~J., {Cochran}, A.~L., {Dello Russo}, N., \& {DiSanti}, M.~A. 2020, \apjl, 889, L10

\bibitem[{{Micheli} {et~al.}(2018){Micheli}, {Farnocchia}, {Meech}, {Buie}, {Hainaut}, {Prialnik}, {Sch{\"o}rghofer}, {Weaver}, {Chodas}, {Kleyna}, {Weryk}, {Wainscoat}, {Ebeling}, {Keane}, {Chambers}, {Koschny}, \& {Petropoulos}}]{Micheli2018}
{Micheli}, M., {Farnocchia}, D., {Meech}, K.~J., {et~al.} 2018, \nat, 559, 223

\bibitem[{{Opitom} {et~al.}(2019){Opitom}, {Fitzsimmons}, {Jehin}, {Moulane}, {Hainaut}, {Meech}, {Yang}, {Snodgrass}, {Micheli}, {Keane}, {Benkhaldoun}, \& {Kleyna}}]{Opitom2019}
{Opitom}, C., {Fitzsimmons}, A., {Jehin}, E., {et~al.} 2019, \aap, 631, L8

\bibitem[{{Opitom} {et~al.}(2025){Opitom}, {Snodgrass}, {Jehin}, {Bannister}, {Bufanda}, {Deam}, {Dorsey}, {Ferrais}, {Hmiddouch}, {Knight}, {Kokotanekova}, {Leicester}, {Marsset}, {Murphy}, {Okoth}, {Ridden-Harper}, {Vander Donckt}, {Ferellec}, {Hutsemekers}, {Lippi}, {Manfroid}, \& {Benkhaldoun}}]{Opitom2025}
{Opitom}, C., {Snodgrass}, C., {Jehin}, E., {et~al.} 2025, arXiv e-prints, arXiv:2507.05226

\bibitem[{{Schleicher}(2010)}]{Schleicher2010AJ}
{Schleicher}, D.~G. 2010, \aj, 140, 973

\bibitem[{{Seligman} {et~al.}(2025){Seligman}, {Micheli}, {Farnocchia}, {Denneau}, {Noonan}, {Santana-Ros}, {Conversi}, {Devog{\`e}le}, {Faggioli}, {Feinstein}, {Fenucci}, {Frincke}, {Hainaut}, {Hoogendam}, {Hsieh}, {Kareta}, {Kelley}, {Lister}, {Mar{\v{c}}eta}, {Meech}, {Oca{\~n}a}, {Pe{\~n}a-Asensio}, {Shappee}, {Taylor}, {Wainscoat}, {Weryk}, {Wray}, {Yaginuma}, {Yang}, \& {Ye}}]{seligman2025arXiv}
{Seligman}, D.~Z., {Micheli}, M., {Farnocchia}, D., {et~al.} 2025, arXiv e-prints, arXiv:2507.02757

\bibitem[{{Seligman} \& {Moro-Mart{\'\i}n}(2022)}]{Seligman2022}
{Seligman}, D.~Z. \& {Moro-Mart{\'\i}n}, A. 2022, Contemporary Physics, 63, 200

\bibitem[{{Tonry} {et~al.}(2012){Tonry}, {Stubbs}, {Lykke}, {Doherty}, {Shivvers}, {Burgett}, {Chambers}, {Hodapp}, {Kaiser}, {Kudritzki}, {Magnier}, {Morgan}, {Price}, \& {Wainscoat}}]{2012ApJ...750...99T}
{Tonry}, J.~L., {Stubbs}, C.~W., {Lykke}, K.~R., {et~al.} 2012, \apj, 750, 99

\bibitem[{{Vernet} {et~al.}(2011){Vernet}, {Dekker}, {D'Odorico}, {Kaper}, {Kjaergaard}, {Hammer}, {Randich}, {Zerbi}, {Groot}, {Hjorth}, {Guinouard}, {Navarro}, {Adolfse}, {Albers}, {Amans}, {Andersen}, {Andersen}, {Binetruy}, {Bristow}, {Castillo}, {Chemla}, {Christensen}, {Conconi}, {Conzelmann}, {Dam}, {de Caprio}, {de Ugarte Postigo}, {Delabre}, {di Marcantonio}, {Downing}, {Elswijk}, {Finger}, {Fischer}, {Flores}, {Fran{\c{c}}ois}, {Goldoni}, {Guglielmi}, {Haigron}, {Hanenburg}, {Hendriks}, {Horrobin}, {Horville}, {Jessen}, {Kerber}, {Kern}, {Kiekebusch}, {Kleszcz}, {Klougart}, {Kragt}, {Larsen}, {Lizon}, {Lucuix}, {Mainieri}, {Manuputy}, {Martayan}, {Mason}, {Mazzoleni}, {Michaelsen}, {Modigliani}, {Moehler}, {M{\o}ller}, {Norup S{\o}rensen}, {N{\o}rregaard}, {P{\'e}roux}, {Patat}, {Pena}, {Pragt}, {Reinero}, {Rigal}, {Riva}, {Roelfsema}, {Royer}, {Sacco}, {Santin}, {Schoenmaker}, {Spano}, {Sweers}, {Ter Horst}, {Tintori}, {Tromp}, {van Dael}, {van der Vliet}, {Venema}, {Vidali}, {Vinther}, {Vola},
  {Winters}, {Wistisen}, {Wulterkens}, \& {Zacchei}}]{Vernet2011}
{Vernet}, J., {Dekker}, H., {D'Odorico}, S., {et~al.} 2011, \aap, 536, A105

\bibitem[{{Ye} {et~al.}(2017){Ye}, {Zhang}, {Kelley}, \& {Brown}}]{Ye2017}
{Ye}, Q.-Z., {Zhang}, Q., {Kelley}, M. S.~P., \& {Brown}, P.~G. 2017, \apjl, 851, L5

\bibitem[{{Zhang} {et~al.}(2022){Zhang}, {Tai}, {Li}, {Zhang}, {Lantz}, {Hiroi}, {Matsuoka}, {Li}, {Lin}, {Wen}, {Han}, \& {Zeng}}]{zhang2022}
{Zhang}, P., {Tai}, K., {Li}, Y., {et~al.} 2022, \aap, 659, A78

\end{thebibliography}

\end{document}